\documentclass[12pt, preprint]{aastex}
\usepackage{natbib}

\slugcomment{Submitted to Astrophys. J.}

\shorttitle{RHESSI Observations of the Solar Flare Fe Line}
\shortauthors{Phillips et al.}

\begin{document}

\title{{\it RHESSI} Observations of the Solar Flare \\Iron-Line Feature at 6.7~keV}

\author{K. J. H. Phillips}\affil{National Research Council Senior Research Associate,
NASA Goddard Space Flight Center,\\
Greenbelt, MD 20771, U.S.A.$^1$}
\altaffiltext{1}{Present address: Mullard Space Science Laboratory,
Holmbury St Mary, Dorking, Surrey RH5 6NT, United
Kingdom}\email{kjhp@mssl.ucl.ac.uk} \and

\author{C. Chifor}\affil{Department of Applied Mathematics and Theoretical Physics, Centre for
Mathematical Sciences, Wilberforce Road, Cambridge CB3 0WA, United
Kingdom} \and
\author{B. R. Dennis}\affil{NASA Goddard Space Flight Center,\\
Greenbelt, MD 20771, U.S.A.}

\begin{abstract}

Analysis of {\it RHESSI} 3--10~keV spectra for 27 solar flares is
reported. This energy range includes thermal free--free and
free--bound continuum and two line features, at $\sim 6.7$~keV and
$\sim 8$~keV, principally due to highly ionized iron (Fe). We used
the continuum and the flux in the so-called Fe-line feature at $\sim
6.7$~keV to derive the electron temperature $T_e$, the emission
measure, and the Fe-line equivalent width as functions of time in
each flare. The Fe/H abundance ratio in each flare is derived from
the Fe-line equivalent width as a function of $T_e$. To minimize
instrumental problems with high count rates and effects associated
with multi-temperature and nonthermal spectral components, spectra
are presented mostly during the flare decay phase, when the emission
measure and temperature were smoothly varying. We found flare Fe/H
abundance ratios that are consistent with the coronal abundance of
Fe (i.e. 4 times the photospheric abundance) to within 20\% for at
least 17 of the 27 flares; for 7 flares, the Fe/H abundance ratio is
possibly higher by up to a factor of 2. We find evidence that the
\ion{Fe}{25} ion fractions are less than the theoretically predicted
values by up to 60\% at $T_e = 12$~MK; the observed $N({\rm Fe
XXV})/N({\rm Fe})$ values appear to be displaced from the most
recent theoretical values by between 1 and 3~MK.

\end{abstract}



\section{Introduction}

The {\it Reuven Ramaty High Energy Solar Spectroscopic Imager} ({\it
RHESSI}) was launched on 2002 February~5 and has since been
returning high-quality X-ray and gamma-ray images and spectra of
solar flares at energies between $\sim 3$~keV and 17 MeV
\citep{den04}. Here we present an analysis of {\it RHESSI}
observations of thermal spectra in the 4--10~keV range for 27
flares. This energy range includes thermal free--free and
free--bound continua and two line features, one at $\sim 6.7$~keV
referred to as the Fe-line feature and the other at $\sim 8$~keV
referred to as the Fe/Ni-line feature. Both features are made up of
lines (resonance lines and satellite lines) due principally to
highly ionized (mostly H-like, He-like, and Li-like) iron (Fe) ions
with the $\sim 8$-keV feature including some lines from highly
ionized nickel. They are both clearly visible in {\it RHESSI} flare
spectra when the plasma temperature is $\gtrsim 20$~MK. The spectral
resolution of the {\it RHESSI} detectors at these energies -- $\sim
1$~keV FWHM -- is inferior to that of crystal spectrometers (e.g.
the Bent Crystal Spectrometer BCS on {\it Yohkoh}), which typically
have an energy resolution of a few eV, enough to resolve the
satellite line structure in the 6.7~keV feature. However, most
crystal spectrometers operating in this range cannot measure the
continuum very accurately since it is overwhelmed by a fluorescence
background. Also, because of their limited spectral coverage, they
have generally not been used to detect the Fe/Ni line feature, apart
from an early observation of \cite{neu69}. {\it RHESSI} covers a
much broader energy range and is able to measure both the two line
features and the continuum emission.

Measurements with {\it RHESSI} of the ratios of the total emission
in the Fe-line feature to the nearby continuum, or more specifically
the equivalent width, are possible. If nonthermal contributions to
the continuum are negligible and if the continuum is emitted by an
isothermal plasma with electron temperature $T_e$ and volume
emission measure $\int_V N_e^2 dV$ ($V$ is the emitting volume and
$N_e$ the electron density), the slope of the logarithm of the
continuum flux with energy $E$ gives $T_e$, and the continuum flux
at a particular energy gives the emission measure. In practice, we
found that an isothermal approximation gives reasonably good fits to
{\it RHESSI} spectra at times after the maximum emission in  soft
X-rays ($< 10$~keV), though not in general at earlier times. This is
discussed further in \S 3. For the 27 flares analyzed here, the
observed equivalent widths of the Fe-line feature as a function of
$T_e$ were compared with the theoretical dependence determined from
ion fractions, line excitation rates, and an assumed value for the
Fe abundance \citep{phi04}. Thus, the Fe abundance in flare plasmas
can, in principle, be determined as a function of time for each
flare. We often found differences between the temperature dependence
of the theoretical and observed equivalent widths, which may be
attributable to incorrect theoretical Fe ion fractional abundances.
This may show a need for the revision of ionization or recombination
rate coefficients for highly ionized Fe ions.

Preliminary results have already been given for a similar analysis
of one flare by \cite{den05}. The flux ratio of the Fe-line feature
to the Fe/Ni line feature is a function of $T_e$ alone, independent
of the iron abundance, and {\it RHESSI} measurements of this ratio
during flares are the subject of a separate paper by \cite{cas06}.

\section{{\it RHESSI} Instrumental Effects}

Instrumental details of {\it RHESSI} are given by \cite{lin02},
\cite{smi02}, and \cite{hur02}. The {\it RHESSI} spectrometer
consists of an array of nine cooled and segmented germanium
detectors. X-ray imaging and spectroscopy up to several hundred keV
are carried out using the 1-cm-thick front segments with the rear
segments operated in anti-coincidence to reduce the background.
Imaging is achieved through the use of rotating modulation
collimators located in front of the detectors, resulting in a
time-modulated signal that can be analyzed to give spatial
information. For 4--10~keV X-rays, the energy resolution (FWHM)
depends on the detector \citep{smi02}, ranging from $\sim 1$~keV for
detectors 1, 3, 4, 5, 6, 8, and 9, $\sim 8$~keV for detector~2, and
$\sim 3$~keV for detector 7. The lower-energy threshold for all
detectors is set at 3~keV except for detectors 2 and 7 for which it
was generally $\gtrsim 10$~keV during the times of the observations
we analyzed. The spectral output from individual detectors or the
summed output of a combination of detectors may be analyzed
independently. For our purposes, the output from detectors 2 and 7
must always be excluded as the lower-energy threshold exceeds the
energies of interest. As indicated below, most of the analysis
reported here was done with the output from detector~4.

Each detector views the flare emission through four beryllium
windows and four blankets of multilayer aluminized-mylar insulation.
The absorption at the X-ray energies of interest here is dominated
by the aluminized mylar. At high photon count rates encountered
during flares, attenuators mounted in front of the detectors move
into place to mitigate pulse pile-up and detector saturation
problems that result from the large ($>10^5$) dynamic range in flare
intensities detectable with {\it RHESSI}. The attenuators consist of
two sets of thin aluminum disks with one set thicker than the other.
Each disk has the same diameter as a detector front segment
(6.15~cm) to attenuate the solar photon flux that reaches the
detector, but has a thinner circular region in the center to allow
some low-energy transmission. Under normal circumstances, both sets
of attenuators are held out of the detector lines of sight to the
Sun. This is referred to as the A0 attenuator state. As the X-ray
counting rate from a flare increases above a prescribed level, the
`thin' attenuators are automatically inserted into the fields of
view. This is called the A1 state. If the emission rises still
further to another prescribed count rate level, then the `thick'
attenuators are also inserted. This is called the A3 state.
Insertion of the thin and thick attenuators results in a reduction
in count rate, particularly at low energies, since the transmission
drops very steeply as a function of decreasing energy. The
transmission fraction drops to 1\% at $\sim 4$~keV in the A0 state,
$\sim 8.5$~keV in the A1 state, and $\sim 13$~keV in the A3 state.
Estimates of the energy-dependent attenuation of the thin and thick
attenuators are available from pre-launch measurements
\citep{smi02}, and are incorporated in the {\it RHESSI} analysis
software package.

The lowest energy photon flux that can be reliably determined in the
different attenuator states depends not only on this increasing
attenuation at the lower energies but also on K-escape processes. An
incident photon with an energy above the germanium K-edge at
11.1~keV can ionize a germanium atom by ejecting one of its two
K-shell electrons. There is a certain probability that the ionized
atom will almost instantaneously emit a K$\alpha$ photon at 9.25~keV
or a K$\beta$ photon at 10.3~keV. These photons, in turn, may be
absorbed in the detector front segment, thus giving a signal
proportional to the full energy of the incident photon.
Alternatively, these secondary photons may escape from the detector
so that the resulting signal is smaller than that expected for the
given incident photon energy. These K-escape events appear in the
count-rate spectrum at an energy equal to the incident photon energy
minus the energy of the escaping K-line photon. Since, in the A1 and
A3 attenuator states, the attenuation at these low energies is very
high, there are very few `true' counts recorded from incident
photons with those energies. In fact, the number of K-escape events
exceeds the number of true events at energies below 5--6~keV and no
information can be obtained about the incident photon spectrum at
these energies. In practice, for reasons at present unknown, there
are usually more counts below 5--6~keV than predicted from our
rather accurate knowledge of the K-escape phenomenon. Consequently,
we have limited our spectral fits for A1 and A3 to higher energies,
i.e. $> 5$~keV, where the predicted and observed count rates can be
made to agree for reasonable model parameters as indicated by the
$\chi^2$ values.

At high photon count rates, spectral distortion takes place because
of ``pulse pile-up." This arises because the instrument electronics
are unable to separate the pulses produced by two photons arriving
in a detector within a few $\mu$s of each other. The result is that
the two photons are recorded as a single event with an energy that
is as high as the sum of the energies of the two photons. In
practice, pulse pile-up begins to appear in the count-rate spectrum
as excess counts at an energy roughly twice the energy of the peak
count rate, i.e. at $\sim 14$~keV in the A0 state, $\sim 20$~keV in
A1, and $\sim 35$~keV in the A3 state. The analysis software allows
some correction for pulse pile-up but with considerably increased
uncertainty in the estimated temperature and emission measure. This
is especially significant at times of increasing count rates
immediately before the attenuators are inserted and times of
decreasing count rates immediately after the attenuators are
removed.

Another rate-dependent effect is a change in the energy calibration
such that the apparent energy of the Fe line at $\sim 6.7$~keV
increases by up to 0.3~keV as the rate increases. This is probably a
fixed increase in the electronics baseline so that the increase is
independent of energy, as is verified by the stability of higher
energy lines in the background spectrum of the front segments. Thus,
it is only significant at lower energies. Nevertheless, it prevents
us at present from exploiting the temperature diagnostic offered by
the slight increase in the Fe-line centroid energy with $T_e$ ($\sim
0.1$~keV over $T_e = 10 - 30$~MK: \cite{phi04}). This effect is
further complicated by the fact that the detector count rates are
rapidly modulated by the collimators as the spacecraft rotates.
Thus, in addition to the energy calibration changing at high rates,
the effective energy resolution is degraded as well.

\begin{figure}
\begin{center}
\includegraphics[width= 14cm,angle=0]{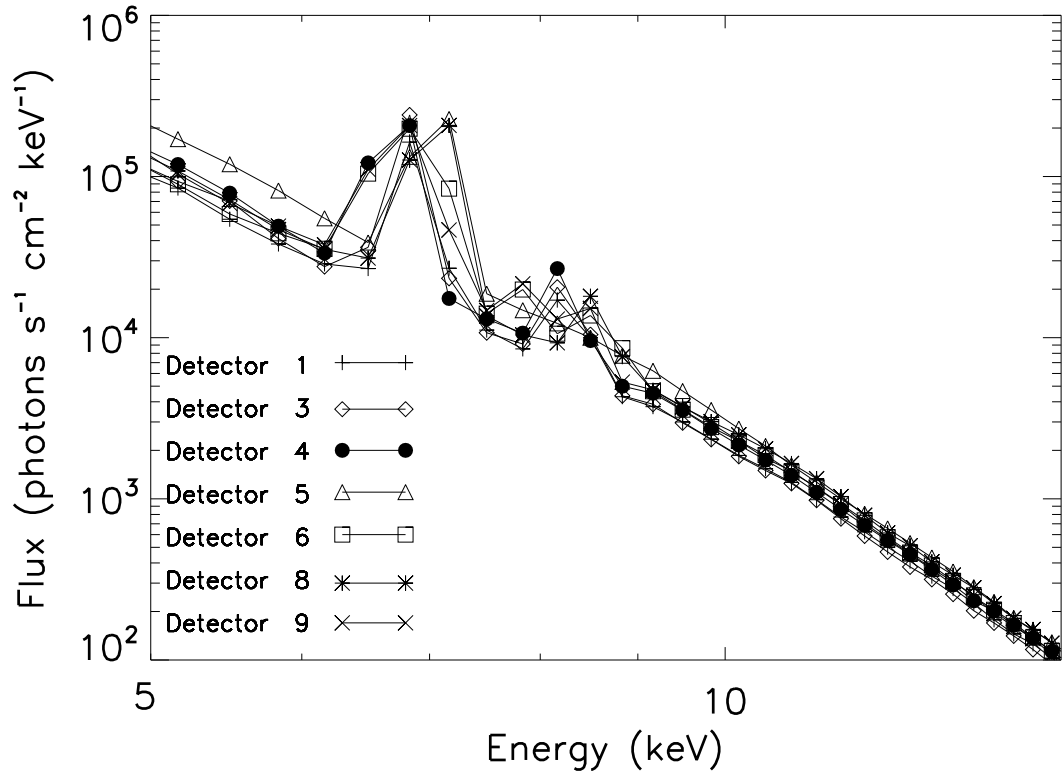}
\caption{Photon spectra obtained from analysis of output from {\it
RHESSI} detectors 1, 3, 4, 5, 6, 8 and 9 over a 20-s interval at
10:01~U.T., during the decay of the flare of 2003 August~19. }
\label{multi_dets}
\end{center}
\end{figure}

With the seven detectors usable for soft X-ray spectral analysis
(all but detectors 2 and 7), we are able to check the measured
sensitivities of each detector by comparing the derived photon
spectra in each case for particular time intervals. In
Fig.~\ref{multi_dets}, we show the comparison of the derived photon
spectra for a time interval when {\it RHESSI} was in the A1 state.
Although the photon fluxes determined from the seven detectors
independently agreed to better than $\sim 20$\% at the higher end of
the energy range chosen for analysis ($\sim 20$~keV), the agreement
at lower energies is less satisfactory. There was typically a factor
2.5 spread from the minimum to maximum derived photon fluxes at
$\sim 5$~keV. This disagreement may be connected with increased
uncertainties in the attenuation at low energies, which is a very
steep function of energy, and with the origin of the excess counts
over and above the K-escape events. Most of the results in the
remainder of this work are from spectra observed with detector~4,
which has relatively good spectral resolution (FWHM $= 0.98$~keV)
and produces a photon spectrum close to the mean of all spectra (see
Fig.~\ref{multi_dets}).

\clearpage
\clearpage

Analysis of {\it RHESSI} flare spectra requires the reliable
determination of background emission \citep{smi02}.  All but 4 of
the 27 flares chosen for analysis here had peak {\it GOES} emission
exceeding M1, so near maximum, the flare soft X-ray count rates were
at least a factor of 100 higher than the background rates at all the
energies used in our fits, i.e. $\lesssim 20$~keV. However, spectra
during periods late in each flare required taking the background
spectrum into careful consideration. The background counts are due
predominantly to the induced radioactivity of the detectors and to
cosmic ray and trapped particle radiation. The rates vary over the
spacecraft orbit, increasing toward higher north and south
geomagnetic latitudes. The rates are also higher after leaving the
South Atlantic Anomaly (SAA) and during passages through
precipitation events at high geomagnetic latitudes. The background
energy spectrum is made up of a continuum and several lines.

\section{Data Analysis}

{\it RHESSI} observations exist of many thousands of solar flares.
We made a selection of flares using whole-Sun X-ray light curves
from {\it GOES} having a reasonably long ($\gtrsim 1$~hour)
duration, particularly during the flare decay. The times of these
flares were then checked against {\it RHESSI} data, removing those
for which {\it RHESSI}\/ data were interrupted by spacecraft night
periods, passages through the SAA, or particle precipitation events.
Having determined that {\it RHESSI} spectra in the A1 attenuator
state gave the most reliable spectral fits (see below), we gave
priority to flares for which the decay stage was mostly in the A1
state. The resulting sample of 27 flares, although not complete by
any particular criterion, does provide enough variety in terms of
flare intensities and {\it RHESSI} coverage for the scope of this
paper.

The {\it RHESSI} data analysis software available in the Solar
Software tree (SSW) requires that the spectral analysis be done in
two stages. In the first stage, the {\it RHESSI} data files
containing the raw time-ordered data covering the times of interest
are read in and count rate spectra (counts s$^{-1}$ per detector)
are extracted over user-supplied time intervals, energy ranges, and
energy bins. For most of the flares analyzed here, detector~4
spectra were extracted for the reasons stated in \S 2. The energy
bins chosen were 1/3~keV wide (the bin width of the telemetered
data) from 3~to~20~keV, and 1~keV wide from 20~to~100~keV.
Corrections are made by the software for detector live time and
energy calibration. Optionally, pulse pile-up corrections can be
made. These are significant if the count rate exceeds $\sim 1000$
counts~s$^{-1}$, and were applied consistently in our analysis.
However, it was found that these corrections were not adequate for
some A0 spectra; when this appeared to be the case, the higher bound
of the energy range used for the spectral fits was taken to be a
value shortward of the affected region.

The output from this first stage of the analysis consists of two
computer files, one containing the count rate spectra for the time
intervals chosen by the user, the other containing the full
spectrometer response matrix (srm) including the off-diagonal
elements that account for photons detected at a lower energy than
their true energy. In the second analysis stage, the two output
files from the first-stage are read by an object-oriented program
known as the Object Spectral Executive ({\sc ospex}). At this stage,
the background is subtracted and instrument-independent photon
spectra are derived from the measured count-rate spectra.

Background subtraction is relatively simple for weaker flares where
the attenuators remain in the A0 state, but becomes more complicated
when the attenuator state changes during the flare of interest. For
flares in the A0 state, short intervals before and/or after the
flare were used to obtain a background spectrum. Linear or low-order
polynomial interpolation between these background intervals in
different energy ranges can give a reliable prediction of the
background during the flare as long as periods of particularly high
background levels are avoided.

The selection of the background spectrum to subtract is usually more
difficult in the A1 and A3 attenuator states, however. This is
because there is usually no time period available either before or
after the flare in the same attenuator state that can be used to
give the background spectrum {\it in that state}. There is often
low-level X-ray emission from somewhere on the Sun both before and
after the flare of interest that is detected in the A0 state but
gives a much lower count rate in the A1 or A3 states. In such cases,
we used the simplest approach of taking the spectrum measured during
the nighttime part of the orbit since then there can be no solar
emission enhancing the background. These background concerns are
only important for times when the estimated background count rates
at any given energy are greater than about 10\% of the measured
count rate from the flare.

With the background spectrum subtracted, the program proceeds
iteratively to find the parameters of a user-selectable model photon
spectrum that best fits the data for each time interval. After
deciding on the components of the model photon spectrum and
reasonable starting parameters, the program uses the spectral
response matrix to calculate the count rate spectrum that would be
produced by the model photon spectrum. This calculated spectrum is
then compared with the measured background-subtracted count-rate
spectrum and a $\chi^2$ value is obtained. The model parameters are
adjusted and the process repeated until a minimum value of $\chi^2$
is obtained. If the value of the reduced $\chi^2$ ($\chi^2$ divided
by the number of degrees of freedom) is $\lesssim 1$, then the
best-fit parameters can be considered to be acceptable for the
particular assumed model photon spectrum.

The model photon spectrum we used included the thermal free-free and
free-bound continuum, a multiple power-law function for the
nonthermal component (when present), and up to three lines with
Gaussian profiles for the Fe-line feature, the Fe/Ni line feature,
and a residual instrumental line at $\sim 10$~keV (see below). In
the version of {\sc ospex} that we used, the model thermal photon
spectra are from the {\sc mekal} \citep{mew85} spectral code with
solar coronal element abundances from \cite{mey85}. For the purposes
of this analysis, an isothermal assumption (single values of $T_e$
and emission measure) was made. This is not likely to be a
particularly close approximation for the rising and peak stages of
most flares, but as we shall see it is a good approximation for many
flares in their decay stages. In the future, we intend to use more
generalized analysis techniques with temperature-dependent (i.e.
differential) emission measures.

We fitted the observed count rate spectra by taking the continuum
function of the {\sc mekal} code with first-guess values for $T_e$
(to define the energy dependence of the continuum) and emission
measure (to define the flux at a particular energy) and with
Gaussian line features at energies of $\sim 6.7$~keV and $\sim
8.0$~keV with first-guess total fluxes to approximate the Fe line
and Fe/Ni line features. These line energies are approximately the
theoretical values for $T_e\gtrsim 20$~MK. The line widths were
generally taken to be 0.1~keV (FWHM) to match the energy range of
the groups of lines forming each feature; the exact widths are
usually irrelevant as they are much less than the spectral FWHM
resolution of $\sim 1$~keV for detector~4. However, at very high
count rates, the line widths were allowed to float also to allow for
the slight degradation of spectral resolution. The goodness of fit
measured by the reduced $\chi^2$ requires estimated uncertainties,
both statistical (i.e. in the count rates) and systematic. The
default value for the systematic uncertainties of 5\% in the
software was used in this analysis; decreasing this value in general
has little effect on the best-fit spectral parameters.

For most of the 27 flares analyzed, we examined spectra over a time
range which generally included the rise phase, the flare maximum,
and the flare decay phase for times when {\it RHESSI} was making
solar observations. Spectra in the A0 state were fitted over an
energy range approximately 4--10~keV, so avoiding any pulse pile-up
peak at $\sim 15$~keV. {\it RHESSI} continuum spectra early in the
flare rise phase in the A0 attenuator state were often poorly fitted
(as shown by the reduced $\chi^2$) with isothermal spectra from the
{\sc mekal} code. The agreement became progressively worse for
higher count rates until the attenuator state changed to A1. There
are three possible reasons for this: (1) the presence of
non-isothermal emitting plasma early in the flare; (2) the
degradation of spectral resolution at high count rates (\S 2); (3)
the worsening goodness of fit at higher count rates owing to the
decrease in statistical uncertainties of the data points. Again, for
flares in the A1 attenuator state, fits to observed spectra in the
rise phase of the flare gave higher $\chi^2$ than fits during the
flare decay, when reduced $\chi^2 \lesssim 1$ could usually be
achieved. The energy range of the A1 spectral fits were from 5.7 or
6.0~keV to 14~keV or more. Because of the strong attenuation below
$\sim 6$~keV and the K-escape events mentioned above, no information
can be obtained about the incident photon spectrum at lower
energies. Many of the results given here are for flares in their
decay stage when {\it RHESSI} was in the A1 state.

In the A3 attenuator state, there was often a poor fit. A
significant reason for this is the appearance of the instrumental
line feature at about 10~keV, so a total of three lines with
Gaussian shapes must be included in the fit. The $\sim 10$~keV line
feature may be due to incorrectly subtracted Ge K-line emission
following radioactive decay of the detector Ge or possibly L$\beta$
line emission from the tungsten collimators in front of detectors
2--9. Part of the line feature at $\sim 8$~keV that we have
attributed to the Fe/Ni line feature may be the tungsten L$\alpha$
line at 8.4~keV. Another factor possibly contributing to the poor
fits obtained in the A3 state is that flares are often near their
maximum at those times. It is likely that the emitting plasma
departs from being isothermal, causing our isothermal model to give
poor fits. The energy range of the fits to A3 spectra was generally
5.7 or 6.0~keV to as high as 30~keV, thus avoiding the pile-up peak
at $\sim 36$~keV.

\begin{figure}
\begin{center}
\includegraphics[width=10cm,angle=90]{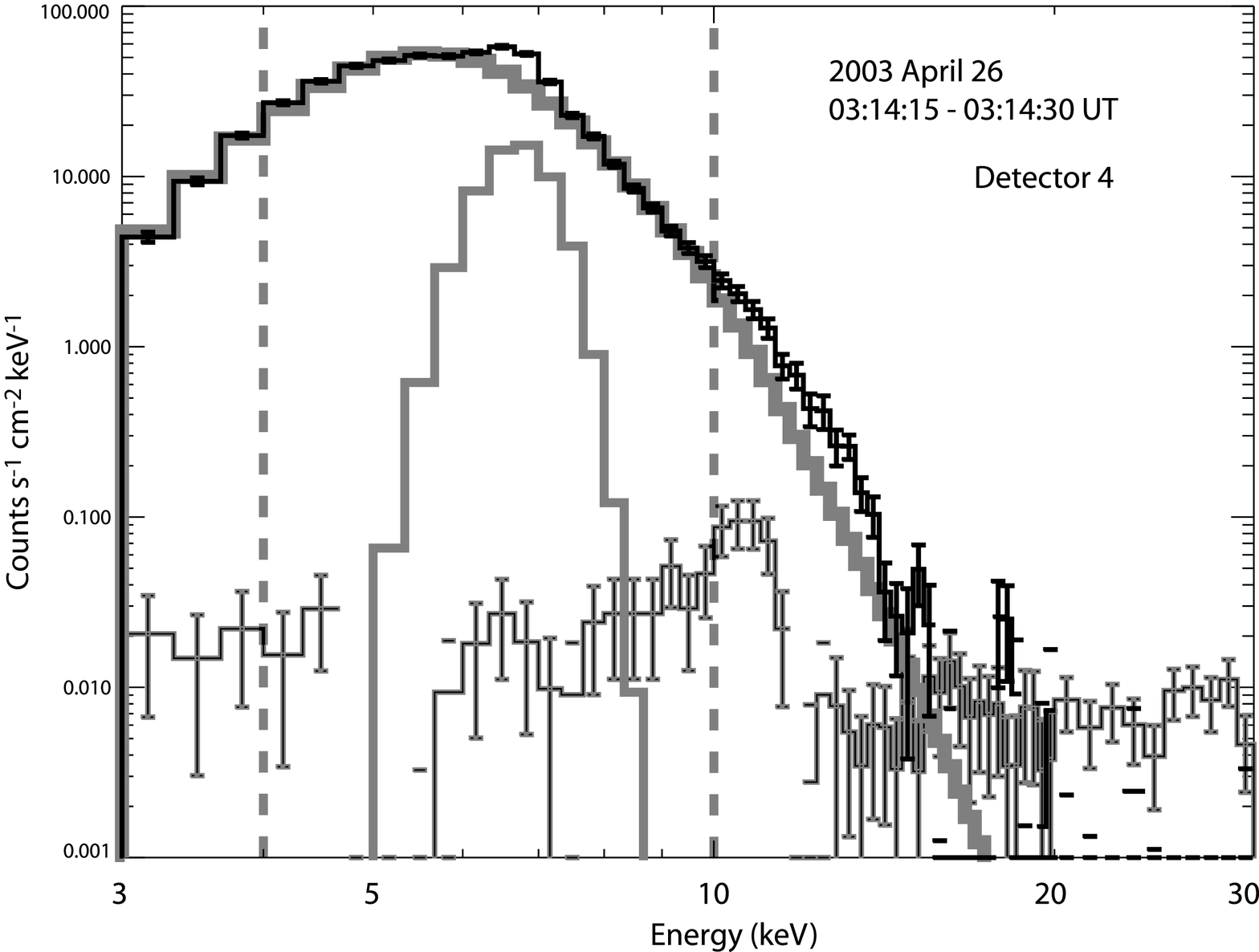}
\includegraphics[width=13.3cm,angle=0]{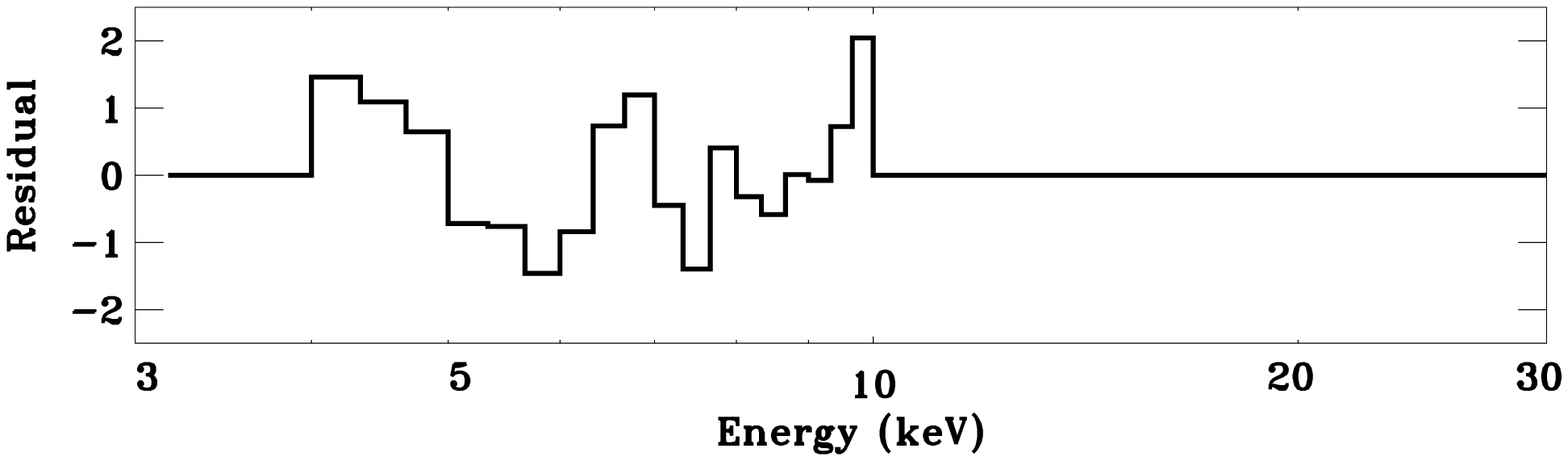}
\caption{{\it Upper panel:} Count rate spectra (counts s$^{-1}$
cm$^{-2}$ keV$^{-1}$) in the energy range 3--30~keV from {\it
RHESSI} detector 4 in the A0 attenuator state. The spectrum is
during the flare on 2003 April 26 (03:14:15--03:14:30 UT). The
histogram with thin dark line and error bars shows the observed
count rate spectrum with uncertainties in each energy bin. The
background spectrum is the histogram with error bars at a count rate
level of $\sim 0.01$. The histograms with thick gray lines show the
MEKAL continuum folded through the spectral response matrix and the
Gaussian line feature representing the Fe-line (energy $\sim
6.7$~keV). The fit range was 4--10~keV (indicated by vertical dashed
lines). The observed excess counts above the model spectrum $>
10$~keV is the result of pulse pile-up and is not included in the
fit. The reduced $\chi^2$ for the goodness of fit was 0.78. {\it
Lower panel:} Residuals (i.e. numbers of standard deviations of the
theoretical count rate above or below the measured rate) in the fit
range. } \label{A0_spectrum}
\end{center}
\end{figure}

\begin{figure}
\begin{center}
\includegraphics[width=10cm,angle=90]{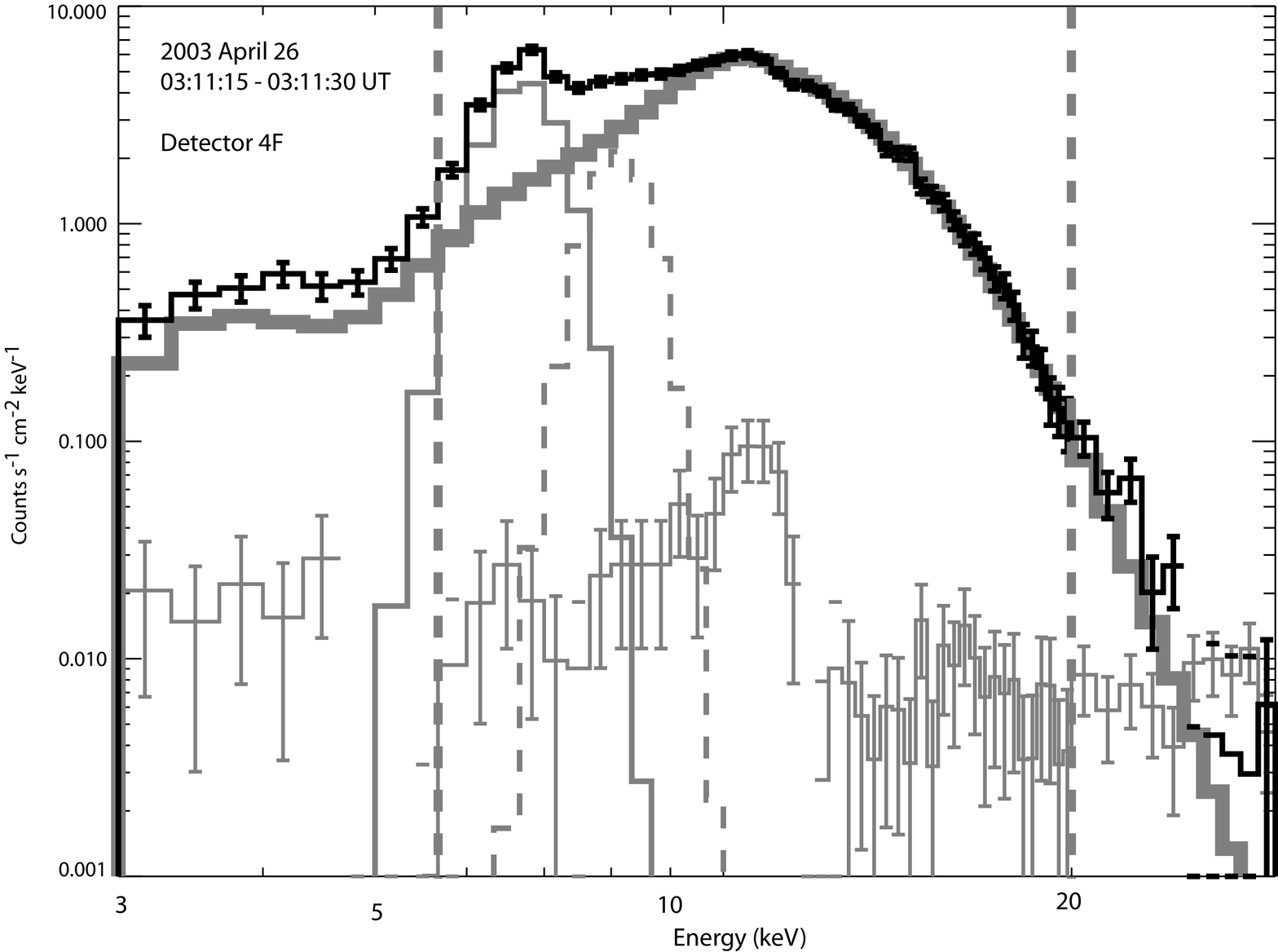}
\includegraphics[width=13.3cm,angle=0]{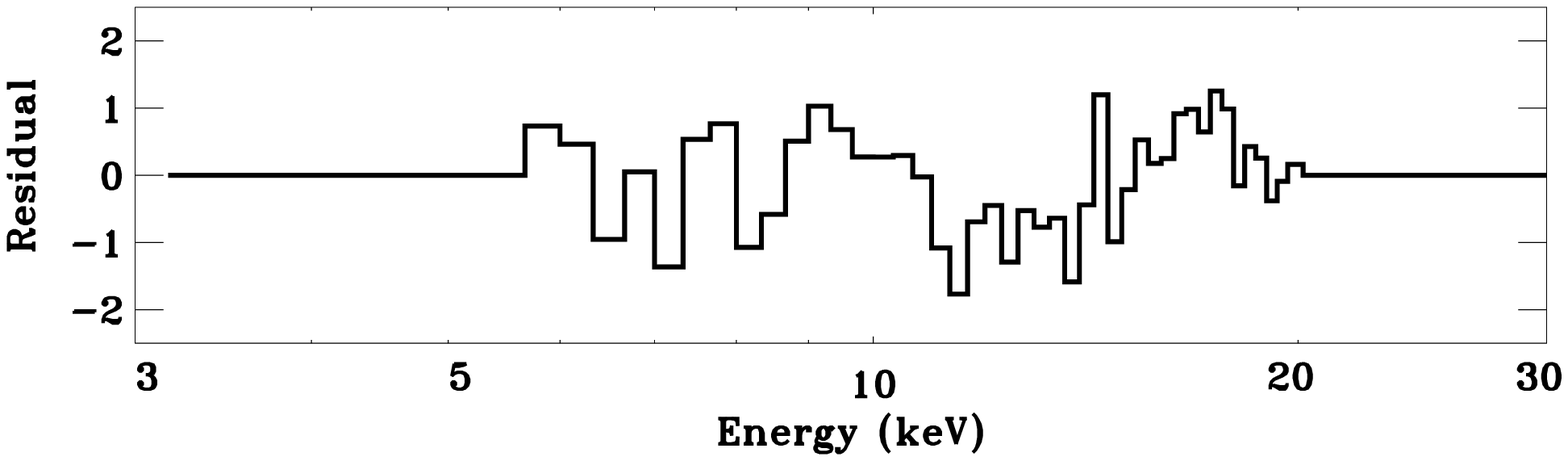}
\caption{ As for Fig.~\ref{A0_spectrum} but for {\it RHESSI}
detector 4 in the A1 attenuator state during the 2003 April~26 flare
(03:11:15--03:11:30 UT). {\it Upper panel:} The model spectrum
includes the Fe line feature as well as a second line feature with
peak energy $\sim 8$~keV (dashed gray histogram) representing the
Fe/Ni line. The fit range indicated by vertical gray lines was
5.7--20~keV. The reduced $\chi^2$ was 0.78. {\it Lower panel:}
Residuals of the fit in the fit range. } \label{A1_spectrum}
\end{center}
\end{figure}

\begin{figure}
\begin{center}
\includegraphics[width=10cm,angle=90]{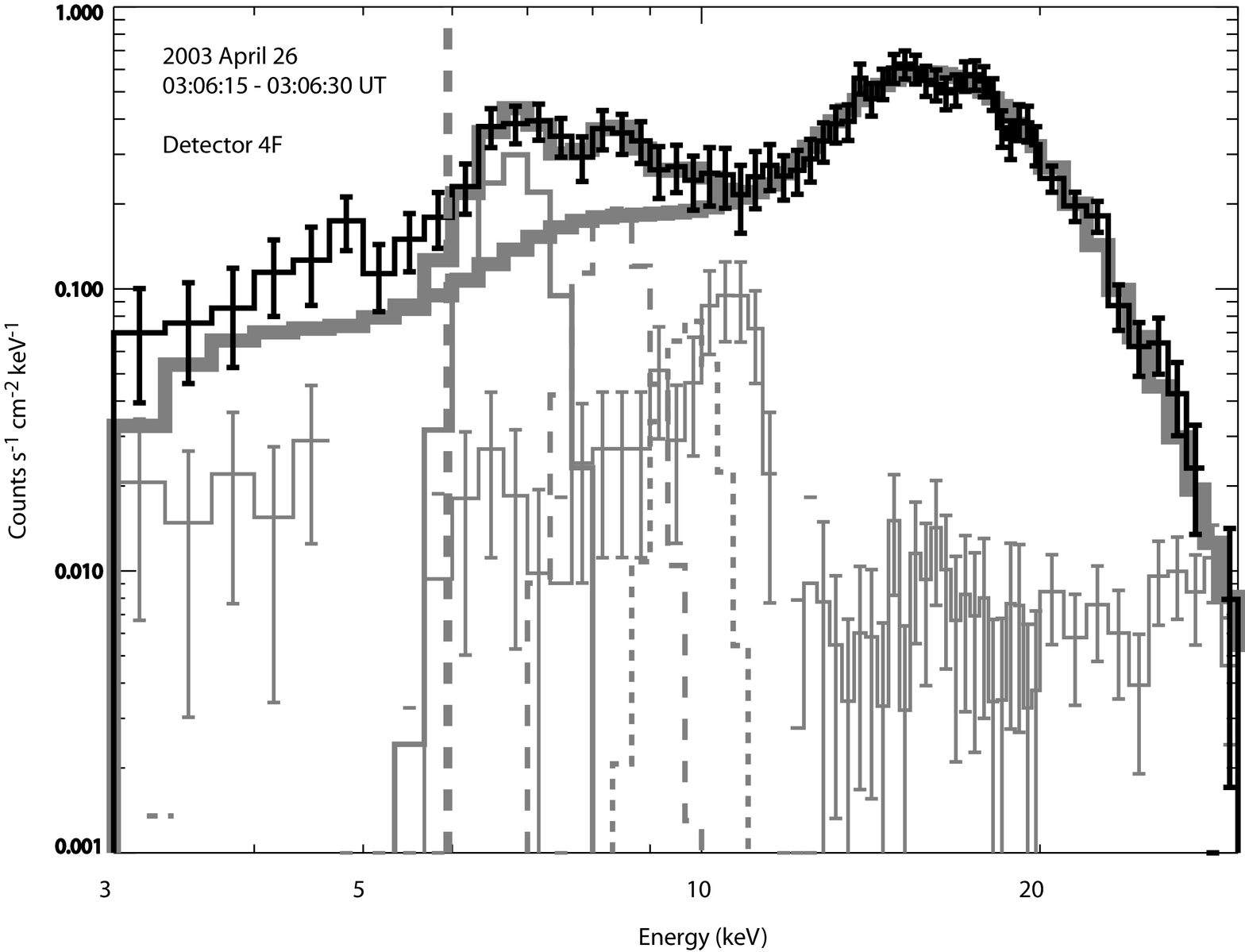}
\includegraphics[width=13.3cm,angle=0]{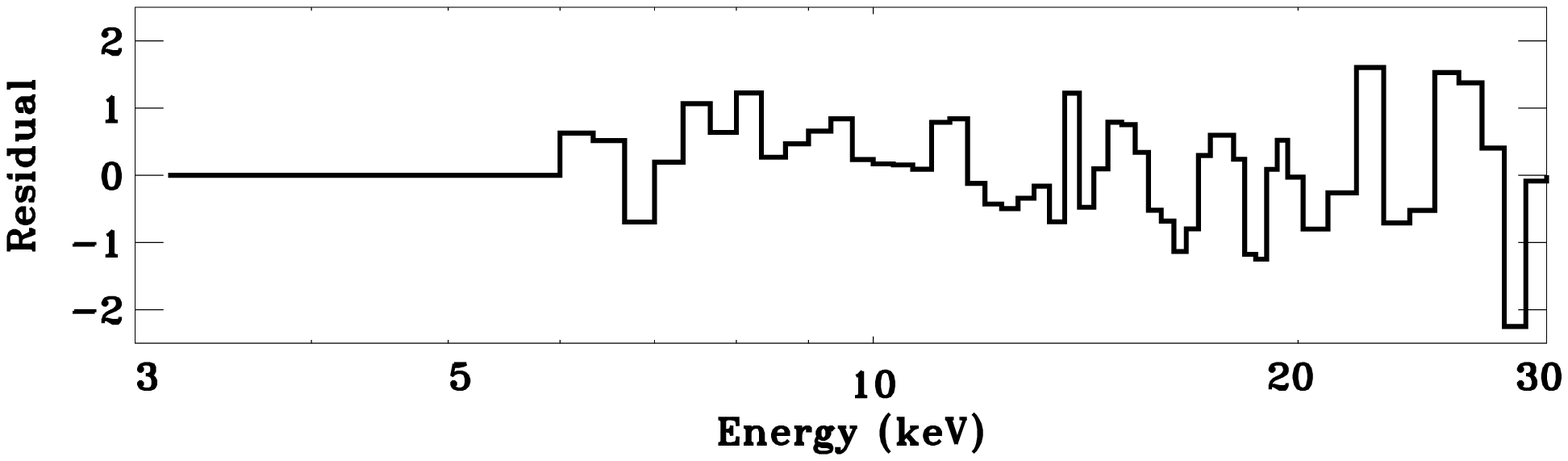}
\caption{ As for Fig.~\ref{A0_spectrum} but for {\it RHESSI}
detector~4 in the A3 attenuator state during the 2003 April~26 flare
(03:06:15--03:06:30~UT). {\it Upper panel:} The Fe and Fe/Ni line
features are indicated as in Fig.~\ref{A1_spectrum}. A third line
feature has been added (gray histogram with short dashes) to account
for an instrumentally formed line at $\sim 10$~keV. The fit range
was 6--30~keV. The reduced $\chi^2$ was 0.69. {\it Lower panel:}
Residuals of the fit in the fit range. } \label{A3_spectrum}
\end{center}
\end{figure}

Figures \ref{A0_spectrum}, \ref{A1_spectrum}, and \ref{A3_spectrum}
show examples of fits to {\it RHESSI} count rate spectra in the A0,
A1, and A3 attenuator states during the M2.1 flare of 2003 April~26.
This flare had two peaks (at 03:05 and 03:10~U.T.), and the X-ray
emission is on the decay of a previous large flare.
Fig.~\ref{A0_spectrum} was taken shortly after the transition to A0
from A1 attenuator states during the decay of the second flare, at
03:14~U.T., with a clear excess of the count rate spectrum over the
model spectrum at $\sim 15$~keV attributable to pulse pile-up
effects and the Fe line feature evident at 6.7~keV. The fitted
spectrum, taken over the range (5--10~keV) to avoid the pulse
pile-up region, includes the {\sc mekal} continuum and Fe line.
Fig.~\ref{A1_spectrum} shows an A1 spectrum during the decay of the
second flare, at 03:11~U.T., with the Fe line feature much more
prominent than in Fig.~\ref{A0_spectrum} and the Fe/Ni line feature
at $\sim 8$~keV appearing as an excess over the continuum on the
high-energy side of the Fe line. The fit range here (5.7--20~keV)
avoids the pulse pile-up region at $\sim 24$~keV. In
Fig.~\ref{A3_spectrum}, the Fe and Fe/Ni line features are stronger
than in Fig.~\ref{A1_spectrum}, with an excess over the continuum at
$\sim 10$~keV attributable to an instrumentally formed line feature.
Note that the observed continuum at low energies ($\lesssim 6$~keV)
in Figs.~\ref{A1_spectrum}  and \ref{A3_spectrum} is poorly
determined owing to the large attenuation in this range and the K
escape events.

\clearpage

\clearpage

For the purposes of this analysis, assumed element abundances affect
the continuum intensities (and therefore emission measure) because
of the free--bound contribution to the continuum. As indicated
above, we used a version of {\sc ospex} that incorporates the {\sc
mekal} code with the solar coronal abundances of \citep{mey85}. Use
of coronal rather than photospheric abundances, for which the
abundances of elements with first ionization potentials $< 10$~eV
are enhanced, is based on the fact that flare plasmas have been
observed to have coronal abundances \citep{flu95, phi03, syl06}. The
abundances of \cite{mey85} differ from the more recent values of
\cite{fel00} by factors of up to 3, but the continuum intensity from
the \cite{mey85} abundances is only 18\% less than that from the
\cite{fel00}  abundances over the 7--12~keV range. If the
\cite{fel00} abundances are more reliable, we may expect that the
emission measures derived in this work are too large by 18\%.

{\it RHESSI} photon spectra have been compared with those derived
from the RESIK crystal spectrometer \citep{syl05} on {\it
CORONAS-F}, which operated between 2001 and 2003. This indicates
that the {\it RHESSI} absolute fluxes at energies $\gtrsim 5$~keV
are accurate to $\sim 20$\%. RESIK is believed to be calibrated in
absolute sensitivity to better than 20\% in its first-order mode
(energy range of 2.0--3.7~keV) with an energy resolution of
$\lesssim 8$~eV or better. Although there is no overlap with the
{\it RHESSI} energy range, the agreement of the extrapolated spectra
is to within the estimated uncertainties of the flux calibration of
both instruments for a period during the 2003 April~26 when {\it
RHESSI} was in its A0 state \citep{den05}. This has been found for
several other flares also, with {\it RHESSI} in A0. On the occasions
when RESIK operated in its third diffraction order, the 6.7-keV
Fe-line feature is included in one of its four channels; RESIK and
RHESSI estimates of the total flux in this line feature during the
2003 April~26 flare agree to better than 50\% \citep{den05}.

\section{Results and Discussion}

\clearpage
\begin{deluxetable}{rclcccccc}
\tabletypesize{\scriptsize} \tablecaption{Analyzed {\it RHESSI}
flares \label{tab1}} \tablewidth{0pt}

\tablehead{\colhead{} & \colhead{Year} & \colhead{Date} &
\colhead{Approx. GOES } & \colhead{GOES} & \colhead{Heliographic} &
\colhead{Approx. UT range }  & \colhead{RHESSI attenuator
 }
 \\ & & & \colhead{U.T. range}& \colhead{class$^a$} &\colhead{coordinates$^a$} &\colhead{of analyzed}  &\colhead{state(s) in the}\\& & &
  \colhead {Start-Peak-End $^a$}& & &\colhead{{\it RHESSI} data}&\colhead{analyzed
UT range}}

\startdata

 1&2002 & Mar. 10/11 & 10/22:21 - 10/23:25 - 11/00:29  &   M2.3&S08  E58  &10/22:57 - 11/01:13 & A1\\
 2& & Mar. 15/16 & 15/22:09 - 15/23:10 - 16/00:42 & M2.2 & S08  W03 & 15/23:33 - 16/00:30 &A1\\
 3& & Apr. 14/15 & 14/23:34 - 15/00:14 - 15/00:25 & M3.7 &N19  W60  &14/23:51 - 15/00:51 &A1\\
 4& & Apr. 21& 00:43 - 01:51 - 02:38  & X1.7& S14  W84 &02:09 - 06:20 &A1 - A0\\
 5& & May. 07& 08:46 - 08:52 - 08:55 &C2.8& S20  W18  &08:52 - 08:57 & A0\\
 6& & May. 31& 00:04 - 00:16 - 00:25 &M2.4& N12  W48  &00:08 - 00:30& A1\\
 7& & Jun. 01&  03:50 - 03:57 - 04:01  & M1.6& S19  E29  &03:53 - 04:03&A1\\
 8& & Jul. 20/21 & 20/21:04 - 20/21:30 -20/21:54 & X3.3 & N17  E72  &20/22:29 - 21/00:38&A1 \\
 9& & Jul. 23 & 00:18 - 00:35 - 00:47 & X4.8 &S13  E72  & 01:00 - 02:30&A1 \\
 10& & Jul. 26/27 & 26/22:36 - 26/22:38 - 26/22:41  & M4.6& S19  E26 & 26/23:01 - 27/00:01&A1 \\
 11& & Jul. 29 & 10:27 - 10:44 - 11:13 & M4.7 &S11  W15  &10:50 - 11:26&A1 \\
 12& & Aug. 24 & 00:49 - 01:12 - 01:31& X3.1 & S02  W81 &01:34 - 02:34&A1 \\
 13& & Oct. 04 & 05:34 - 05:38 - 05:41 & M4.0& S18  W08 &05:41 - 05:48&A1 \\
 14& & Dec. 02 & 19:19 - 19:27 - 19:33& C9.6& N09  E32 &19:26 - 19:32&A1 \\
 15& & Dec. 17 & 22:57 - 23:35 - 23:45 & M1.6& S27  E01 &23:20 - 23:52&A0 - A1 \\
 16& 2003 & Apr. 23 &  00:57 - 01:06 - 01:30 &   M5.2& N18  W16 &01:02 - 01:17 & A1\\
  17 & & Apr. 26 &  03:01 - 03:06 - 03:12 &   M2.1& N19  W66  &03:04 - 03:29 & A3 - A1 - A0\\
  18& & May. 29 &  00:51 - 01:05 -  01:12  &   X1.1& S07  W31  &00:45 - 01:21 & A0 - A1\\
  19& &Aug. 14 &  22:28 - 22:44 - 23:15 &   C5.0&S17  W87 & 22:40 - 22:52 & A1 - A0\\
  20& &Aug. 19 & 09:45 - 10:06 - 10:25  &   M2.7& S15  W68 & 09:59 - 10:45 & A1\\
  21& &Oct. 22 &  19:47 - 20:07 - 20:28 &   M9.9& S17  E88  &20:16 - 20:37 & A1\\
  22& &Oct. 23 & 19:50 - 20:04 - 20:14  &   X1.1& S17  E88 &19:59 - 20:31 & A3 - A1\\
  23& &Nov. 02 & 17:03 - 17:25 - 17:39  &   X8.3&S18  W59 & 18:37 - 19:36 & A1\\
   24& &Nov. 11 &  15:23 - 16:15 - 17:17 &   C8.5& S00  E90 &16:00 - 18:10 & A1 - A0\\
 25& 2004  &Jan. 05 & 02:50 - 03:45 - 05:20  &   M6.9& S10  E36 &04:06 - 04:52 & A1\\
  26& &Jul. 20 & 12:22 - 12:32 - 12:45  &   M8.7& N10  E33  &12:37 - 12:47 & A3 - A1\\
 27& 2005
   &Jan. 15/16 & 15/22:25 - 15/23:02 - 15/23:31  &   X2.6& N14   W08  & 16/01:30 - 16/03:13 & A1\\

\enddata


\tablenotetext{a} {Data from NOAA \textit{Solar X-ray Flares} and
the Lockheed Martin Solar and Astrophysics Laboratory \textit{Latest
Events Archive}}

\end{deluxetable}
\clearpage

Table~\ref{tab1} gives details of the 27 flares between 2002 and
2005 analyzed in this work. The {\it GOES} X-ray importance of these
flares ranges from C2.8 to X8.3. The times of {\it RHESSI} spectral
analysis and the attenuator states are given, as are the flare
locations on the Sun.

\begin{figure}
\begin{center}
\includegraphics[width= 12cm,angle=0]{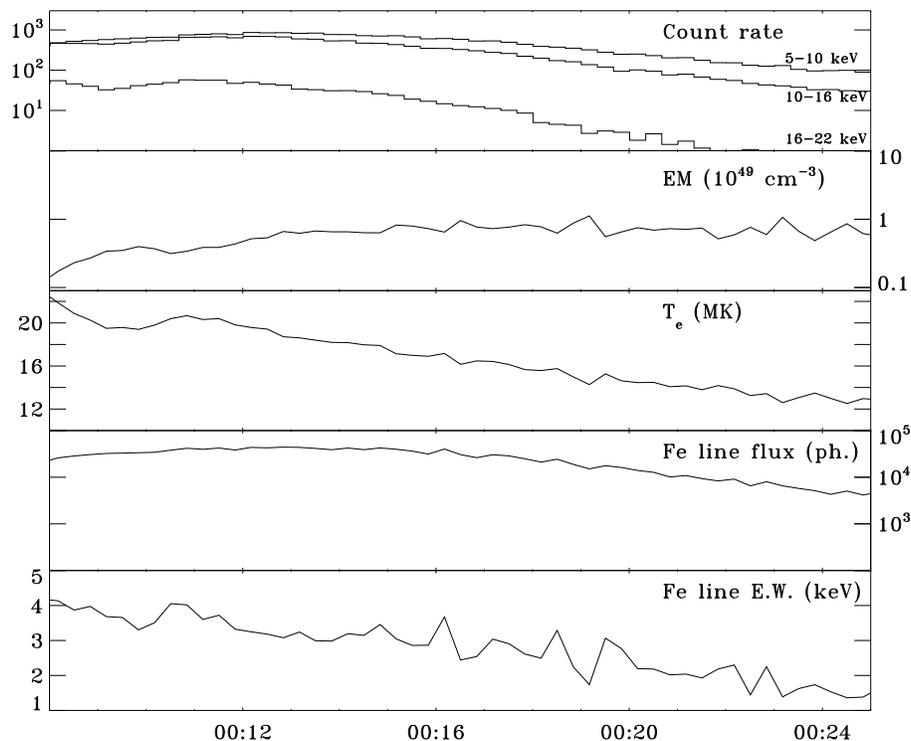}
\caption{{\it RHESSI} 5-10 keV, 10-16~keV and 16-22~keV count rates
(s$^{-1}$), emission measure (EM, units of $10^{49}$ cm$^{-3}$,
logarithmic scale) and $T_e$ (MK), Fe-line feature flux (`ph.' =
photons cm$^{-2}$ s$^{-1}$) and equivalent width (keV) for the M2
flare on 2002 May~31 (No.~6 in Table~\ref{tab1}). The peak of the
flare in {\it GOES} was at 00:16~U.T. The background spectrum in a
40-s night-time interval before the flare, at approximately
23:55~U.T., was subtracted from the measured count rate spectrum in
each time interval during the flare. The uncertainties in these
plots are indicated by the scatter of points about smooth curves
through the points in each case. } \label{May31_timeplots}
\end{center}
\end{figure}

Figure \ref{May31_timeplots} shows the {\it RHESSI} count rates in
three energy bands (background-subtracted), as well as the time
histories of estimated values of emission measure, $T_e$, and the
flux and equivalent width of the Fe-line feature during the M2 flare
of 2002 May~31 (No.~6 in Table~\ref{tab1}). Over this interval as
well as the flare peak, {\it RHESSI} was in the A1 state. The photon
counts in the 16--22~keV range indicate that there was no
significant nonthermal component, a short impulsive phase having
occurred at about 23:57~U.T. on May~30. A fit to 68 spectra over the
range 5.3--16.7~keV in 20-s intervals shows close agreement between
observed and model spectra having the {\sc mekal} continua and two
line features to match the Fe and Fe/Ni line features, with reduced
$\chi^2$ steadily decreasing from approximately 1.2 at the earlier
times to 0.8 at later times. An isothermal emitting plasma is a good
approximation for all spectra, particularly those over the decline
of this flare.

\clearpage
\begin{figure}
\begin{center}
\includegraphics[width= 12cm,angle=0]{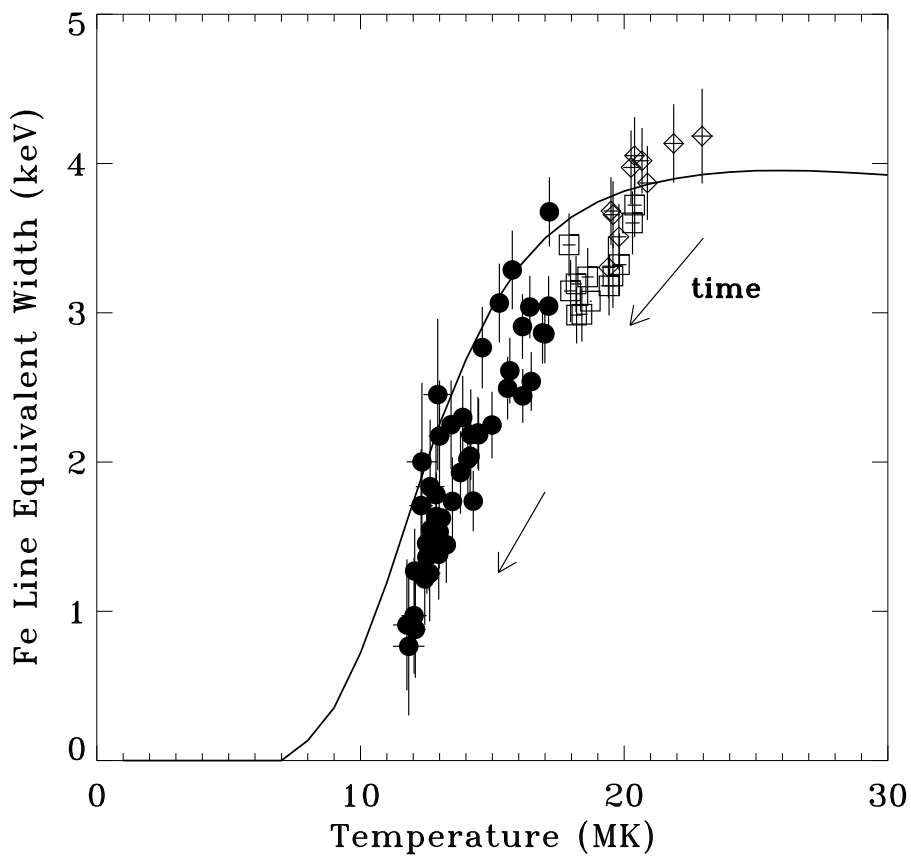}
\caption{Measured Fe line equivalent widths (keV) plotted against
$T_e$ (MK) during the flare of 2002 May 31 (No.~6 in
Table~\ref{tab1}). The solid curve is the theoretical dependence,
revised from the calculations of \cite{phi04}. Points in the A1
state during the rise phase (00:08--00:11~U.T.) of this flare are
shown as diamonds; those at the peak (00:11--00:16~U.T.) as squares;
those in the decay (00:16--00:30~U.T.) as black circles. The general
sequence in time is indicated by arrows. } \label{May31_2002_EQW}
\end{center}
\end{figure}

\clearpage The estimated equivalent widths of the Fe-line feature
were plotted against $T_e$ for all intervals and compared with the
theoretical values, taken from an updated version of that given by
\cite{phi04}. The theoretical curve has been revised as a result of
changed definitions of the Fe line flux and the continuum flux at
the line energy. In \cite{phi04}, the line flux was taken to be the
excess over the continuum after smoothing calculated spectra from
the CHIANTI code to 1-keV resolution; this is here redefined to be
the total of lines making up the Fe-line feature without smoothing.
Also, in \cite{phi04}, the line energy was taken to be the energy of
the centroid of the line with continuum emission included, whereas a
more satisfactory definition is the centroid energy of the line
alone. As in \cite{phi04}, Feldman \& Laming's (2000) coronal
abundance of Fe (Fe/H $= 1.26\times 10^{-4}$) is assumed, together
with the ionization fractions of \cite{maz98}. The updated
definition of the Fe line equivalent width leads to a maximum
theoretical value of 4.0~keV at $T_e \sim 24$~MK, compared with
3.0~keV at $T_e \sim 23$~MK \citep{phi04}.
Figure~\ref{May31_2002_EQW} is a plot of the Fe line equivalent
width, including the updated theoretical curve and observed points
for the 2002 May 31 flare, the time variations of which are shown in
Fig.~\ref{May31_timeplots}. The observed equivalent widths were
estimated from spectra in the A1 attenuator state over the period
from shortly before the flare maximum to about 10 minutes after.
From Fig.~\ref{May31_2002_EQW}, it can be seen that the observed
points follow a dependence with temperature that approximates the
theoretical curve. The points on the flare decay are nearer to the
curve than those on the flare rise or at the flare peak. Most points
are generally lower than and to the right of the curve.

\clearpage
\clearpage

\begin{figure}
\plotone{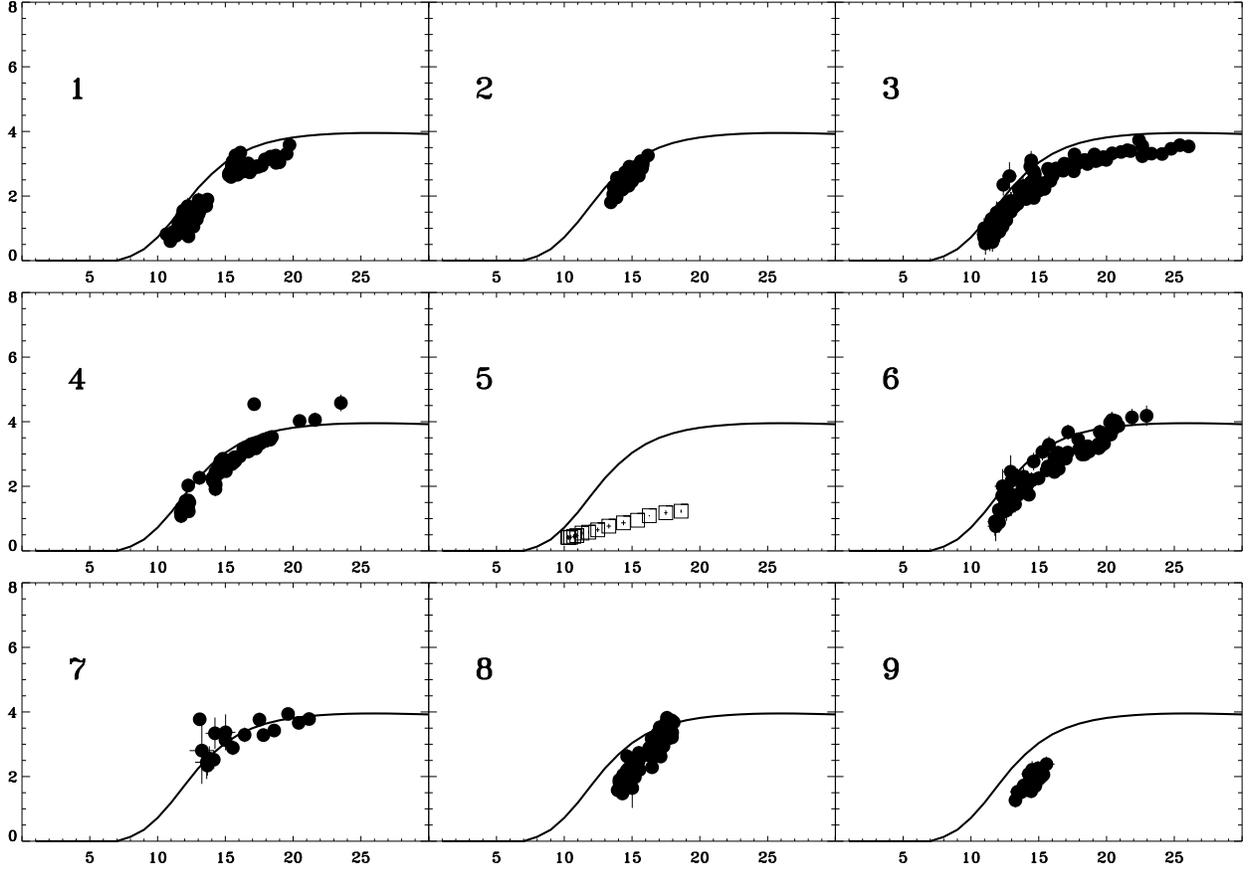} \caption{Fe-line feature equivalent widths (keV,
vertical axis) plotted against $T_e$ (MK, horizontal axis) estimated
from {\it RHESSI} spectra during the 27 flares listed in
Table~\ref{tab1}. The curve is the theoretical equivalent width vs.
$T_e$ for a coronal abundance of Fe (Fe/H $= 1.26\times 10^{-4}$)
and ionization fractions of \cite{maz98}. In this figure, estimated
equivalent widths are distinguished according to the {\it RHESSI}
attenuator state: squares (A0), circles (A1), and triangles
(A3).}\label{Eqw_plots}
\end{figure}

Plots for other flares similar to Fig.~\ref{May31_2002_EQW} show
that the equivalent width increases with $T_e$ in ways that have
varying degrees of agreement with the theoretical curve.
Figure~\ref{Eqw_plots} shows Fe line equivalent width plots against
$T_e$ for each of the 27 flares of this analysis. The theoretical
curve identical to that in Fig.~\ref{May31_2002_EQW} is shown for
each flare. As indicated in Table~\ref{tab1}, most estimates of Fe
line flux and equivalent width were made when {\it RHESSI} was in
its A1 state since it was then that the best $\chi^2$ values were
achieved. Estimates in other attenuator states are indicated in
Fig.~\ref{Eqw_plots} by different symbols (see figure caption).

Patterns of the observed equivalent width variations can be
discerned for most flares. These are most obvious for the 20 flares
(nos. 1, 3, 4, 5, 6, 7, 10, 12, 13, 14, 15, 16, 17, 18, 19, 20, 21,
22, 25, and 26) for which there is a sequence of measurements
covering a relatively large temperature range ($\gtrsim 5$~MK),
often well into the flare decay. Most or all of the equivalent width
estimates for each flare were made in the A1 attenuator state with
the exceptions of flare~5, made in the A0 attenuator state, and
flares~19, 22, and 26, partly made in the A3 attenuator state.  For
four flares (1, 3, 12, and 17), the observed equivalent widths lie
below the theoretical curve, with clear evidence in at least the
case of flare~3 that the observed equivalent widths saturate at high
temperatures ($T_e \gtrsim 20-25$~MK). The observed maximum
equivalent width in these four flares is approximately 80--90\% of
the theoretical maximum equivalent width (4.0~keV), with the
observed equivalent widths at lower temperatures less than 80\% of
the theoretical value. For flare~5 ({\it GOES} class C2.8), the
equivalent widths are only 40\% of the theoretical values, although
the A0 spectra at higher temperatures are fitted poorly by the model
spectra and may be unreliable. For flare~7, the observed points
cluster around the theoretical curve, while those for flare~20 are
below the theoretical curve for $T_e \lesssim 18$~MK and agree with
it for $T_e \gtrsim 18$~MK.

The equivalent widths of at least 15 flares (2, 4, 6, 8, 10, 13, 14,
15, 16, 18, 21, 22, 25, 26, and 27) rise with $T_e$, crossing the
theoretical curve at temperatures from 15~MK to 28~MK. For flares~22
and 26, there is some evidence that without the A3 points the
maximum equivalent width would be only slightly greater than 4~keV.
In view of the often poor fits to model spectra with {\it RHESSI} in
its A3 state, the presence of some A3 data may be misleading. For
each of these 15 flares, the observed points rise with $T_e$
following a path that is clearly to the right of the theoretical
curve.

\notetoeditor{The three parts of Figure 7 should be printed over
three pages in published edition of this article. }
\clearpage {\plotone{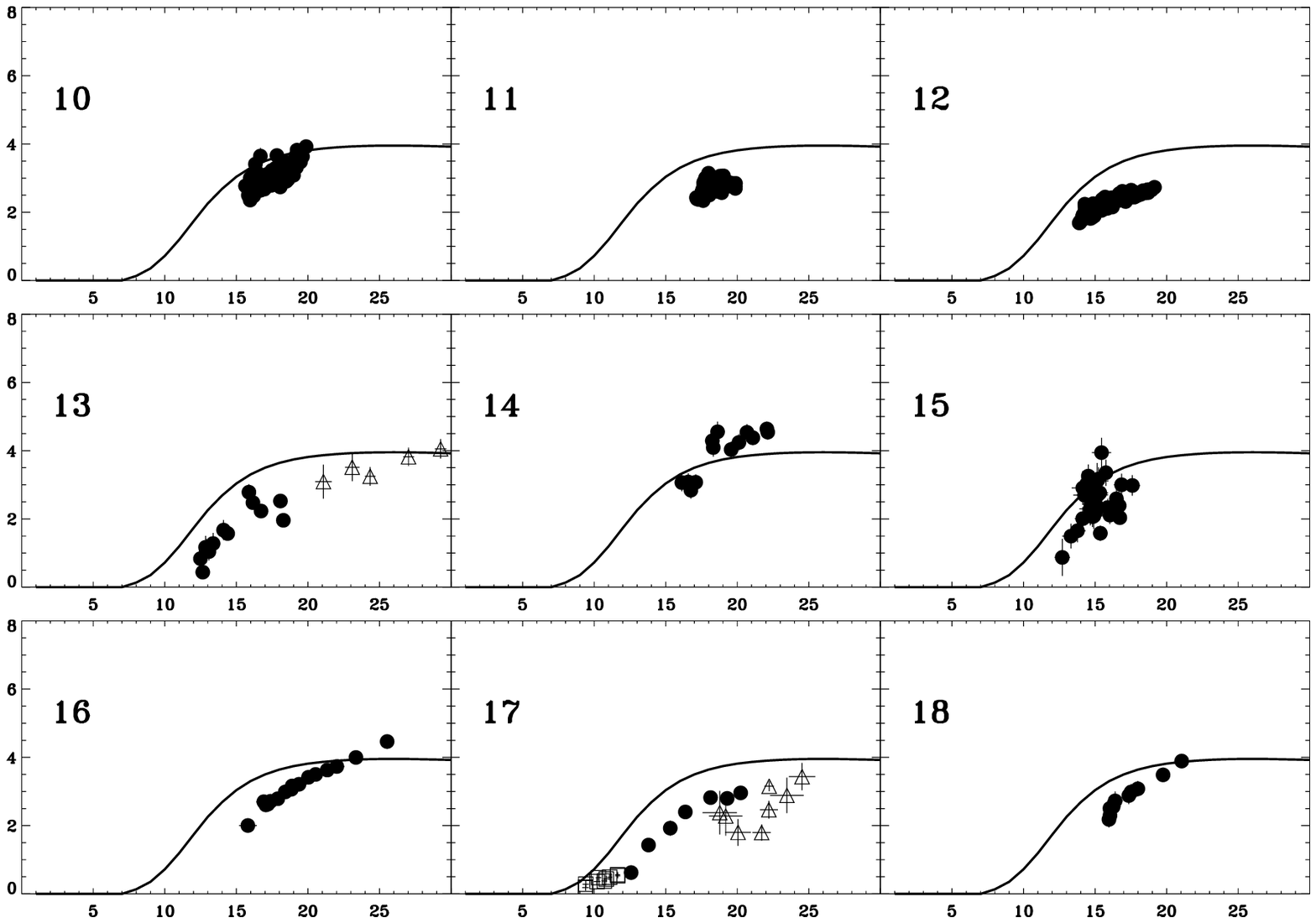}} \centerline{Fig. 7. --- Continued. }
\clearpage {\plotone{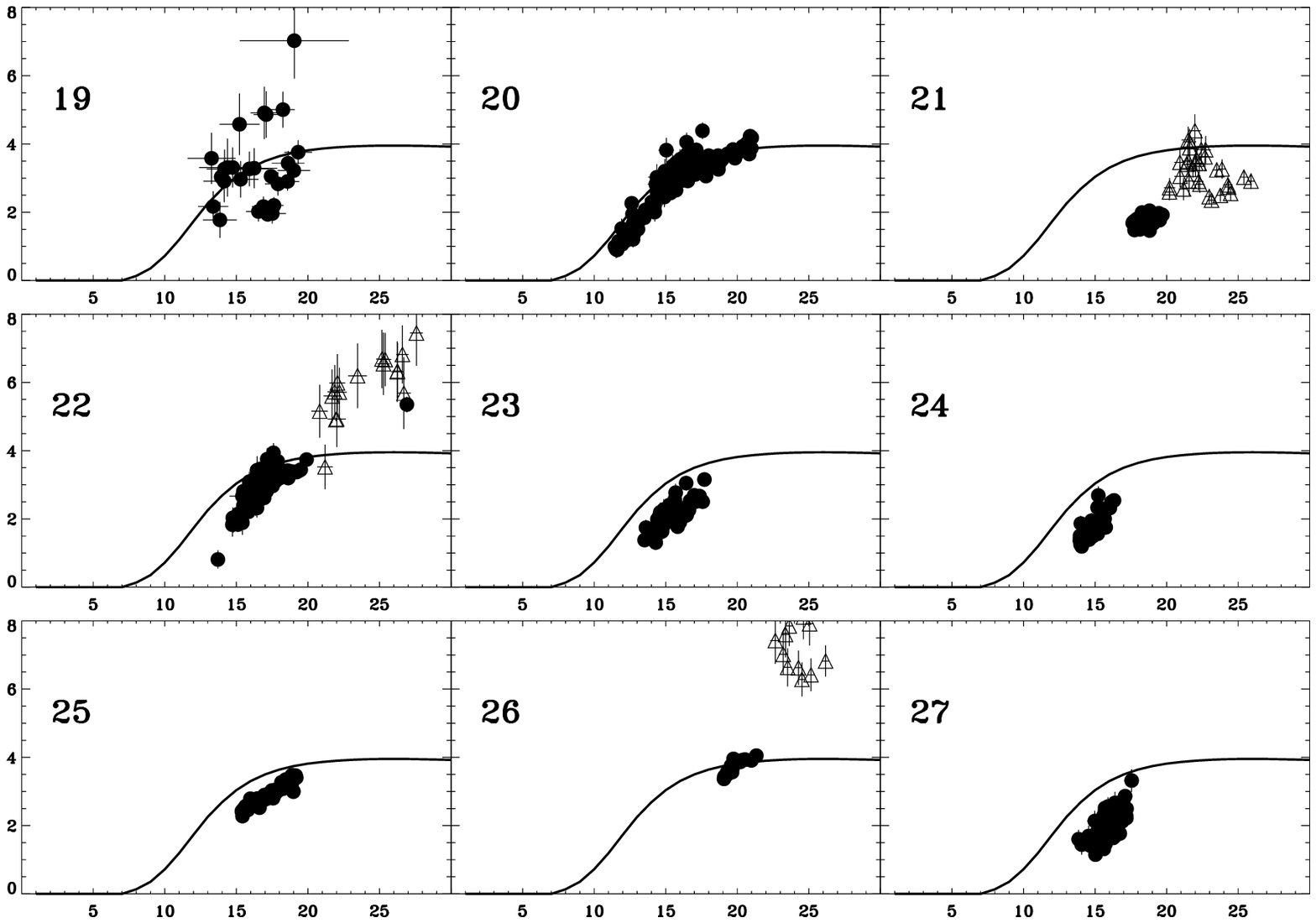}} \centerline{Fig. 7. --- Continued. }
\clearpage
\section{Summary and Conclusions}

For this analysis of {\it RHESSI} solar flare data, low-energy
($\gtrsim 5$~keV) spectra during flares were fitted with model
thermal spectra consisting of a continuum (free--free plus
free--bound emission) from the {\sc mekal} spectral code and line
features with Gaussian profiles at 6.7~keV (the Fe line), $\sim
8$~keV (the Fe/Ni line), plus an instrumentally formed line at $\sim
10$~keV. The data analysis was performed with software that accounts
for all known instrumental effects and allows for background
subtraction. An isothermal approximation was used, so that the
continuum slope and flux measure the temperature $T_e$ and emission
measure $\int_V N_e^2 dV$. The fits to the model spectra are most
reliable (as measured by the reduced $\chi^2$) for {\it RHESSI}
spectra taken in its A1 attenuator state during the decay of flares.
Some 27 flares were selected from the {\it RHESSI} data set, chosen
for their relatively long durations so that repeated measurements of
the Fe line equivalent could be made, particularly during their
decay stages. Measurements of the Fe line equivalent width were made
from these fits to the observed spectra, and plots made for each
flare of the equivalent width against $T_e$. Comparison was made
with an updated version of the theoretical curve from \cite{phi04},
which assumes the coronal abundance of Fe (Fe/H $= 1.26\times
10^{-4}$, i.e. $4\times$ photospheric) derived by \cite{fel00}. For
only one flare (no. 7) is there good agreement between theory and
observation, and near-agreement for another flare (no. 20). For four
other flares (1, 3, 12, 17), the observed equivalent widths lie
below the theoretical curve, the observed values reaching a maximum
which is 80\% of the theoretical value, $\sim 4$~keV. Agreement
between the observations and theory for these four flares could be
achieved if the coronal relative abundance of Fe were multiplied by
0.8 (i.e. Fe/H $\sim 1 \times 10^{-4}$ or $3.2 \times$
photospheric).

A more common characteristic is for the observed equivalent widths
to rise with $T_e$, exceeding the theoretical curve at $T_e = 15 -
28$~MK. With A3 spectral results included, the equivalent width for
a few flares rises to a value of $\sim 8$~keV, more than double the
maximum of the theoretical curve. If the A3 points are disregarded,
the observed points rise to a value of slightly more than 4~keV. The
observed equivalent widths increase with $T_e$ more sharply than is
predicted by the theoretical curve, with a clear displacement to
higher temperatures being indicated. These results are not so easily
explained. If the isothermal approximation is a good one for spectra
during these flares, as is indicated by the reduced $\chi^2$ values,
then possibly this behavior could indicate the need for a correction
to the ionization fractions of \cite{maz98}. In particular, the
displacement indicates that the value of $N({\rm Fe}^{+24})/N({\rm
Fe})$ ($N({\rm Fe}^{+24}) =$ the number density of He-like Fe and
$N({\rm Fe}) =$ the number density of all Fe ions) should be lower
than is calculated for the temperature range $12<T_e\, ({\rm
MK})<18$.

\cite{ant87} suggested corrections to the $N({\rm Fe}^{+23})/N({\rm
Fe}^{+24})$ ratio based on {\it Solar Maximum Mission} crystal
spectrometer data. This ratio is equal to $Q/R$ where $Q$ is the
rate coefficient of ionizations from Fe$^{+23}$ and $R$ the rate
coefficient of recombinations to Fe$^{+24}$, both functions of
$T_e$. \cite{ant87} argued that, since the various theoretical rate
coefficients of ionization from the Fe$^{+23}$ available then had
more scatter than the rate coefficients of recombinations to
Fe$^{+24}$, and that the recombination rates were more slowly
varying with $T_e$ than the ionization rates, the ionization rates
were more likely to require revision than the recombination rates.
The ionization rate coefficients in the work of \cite{maz98} are
based on analytical formulae that are known to be a good
representation of measured data for ions up to Fe$^{+15}$
\citep{arn92}. However, there is a need for the inclusion of
ionization fractions based on experimental data for ionization rates
from Fe$^{+23}$ and other ions, now available (e.g. \cite{won93}).
Our result is consistent with the number density of Fe$^{+24}$ ions
being too small compared with that of lower stages like Fe$^{+23}$,
but it is difficult to explain why the temperature displacement
varies from flare to flare, as found here. Possibly non-equilibrium
effects play a role, but this is unlikely in view of the probable
high densities ($N_e \sim 10^{11}$ cm$^{-3}$) of the emitting
plasma, as was shown by \cite{phi04}.

The tendency of Fe-line feature equivalent widths to agree better
with theory for periods when flares are in their declining stage and
{\it RHESSI} is in its A1 attenuator state is probably due to the
flares having a more nearly isothermal state in their decay. This
has also been found from temperature measurements using \ion{Ca}{19}
and \ion{S}{15} line ratios using the Bragg Crystal Spectrometer on
the {\it Yohkoh} spacecraft \citep{phi05b}. Flares probably depart
most strongly from an isothermal state in their early and peak
stages, when images (particularly those from {\it RHESSI} and the
{\it TRACE} 195~\AA\ filter: \cite{gal02}, \cite{phi05a}) show
emission in loop-top structures at high temperatures in close
proximity with loop structures with much lower temperatures. The
general lack of agreement of Fe line feature equivalent widths
observed late in flare developments by {\it RHESSI} in the A0 state
is therefore unexpected on this basis, but this appears to be due,
at least occasionally, to the degraded spectral resolution at high
count rates and the difficulty of distinguishing the line feature at
low count rates and at low temperatures.

Multi-thermal flare plasmas are being investigated and will be the
subject of a further analysis of {\it RHESSI} flare spectra.
Differential emission measures, DEM $= N_e^2 dV/dT_e$, can be
extracted from spectral line fluxes and other data. Procedures like
the PintofAle code \citep{kas00} and DEMON \citep{syl80} are
currently being compared, using spectral data from {\it RHESSI}, the
{\it CORONAS-F}  RESIK crystal spectrometer, {\it GOES}, and other
broad-band data. Simpler procedures using analytical forms for DEM
are also being tried, such as DEM $=$ exp$(-T_e/T_0)$ ($T_0 $ is a
constant for a particular time), since RESIK data for \ion{Si}{12}
dielectronic satellite line intensity ratios show that such a
function is an improvement over an isothermal approximation
\citep{phi06}.

K.J.H.P. acknowledges support from a National Research Council
Senior Research Associateship award. C.C.'s work was supported by a
Research Assistantship from the Catholic University of America and
NASA/GSFC. We are grateful to M. Berg, D. W. Savin, R. A. Schwartz,
and A. K. Tolbert for their help.

\end{document}